\documentclass[letterpaper,12pt]{article}
\usepackage{tabularx} 
\usepackage{amsmath}  
\usepackage{amssymb}
\usepackage[pdftex]{graphicx}
\pdfoutput=1
\usepackage[margin=1in,letterpaper]{geometry} 
\usepackage{cite} 
\usepackage{authblk}
\usepackage{color}

\newcommand{\beq}{\begin{equation}}
\newcommand{\eeq}{\end{equation}}
\newcommand{\bea}{\begin{eqnarray}}
\newcommand{\eea}{\end{eqnarray}}

\newcommand{\ba}{\begin{array}}
\newcommand{\ea}{\end{array}}

\begin{document}

\setlength{\baselineskip}{0.7cm}
\begin{titlepage}
    \begin{flushright}
        NITEP 141\\
        KEK-TH-2436
    \end{flushright}
    \vspace{10mm}
    \begin{center}
        \Large\textbf{Gauge Coupling Unification} \\
        \vspace{2mm}
        \Large\textbf{in simplified Grand Gauge-Higgs Unification} \\
    \end{center}
    \vspace{10mm}
    \begin{center}
        {\large Nobuhito Maru},$^{a,\,b}$
        {\large Haruki Takahashi}$^{\,c,\,d}$ and
        {\large Yoshiki Yatagai}$^{\,a}$
    \end{center}
    \vspace{0.2cm}
    \begin{center}
        ${}^{a}$ \textit{Department of Physics, Osaka Metropolitan University, \\
        Osaka 558-8585, Japan} \\
        ${}^{b}$ \textit{Nambu Yoichiro Institute of Theoretical and Experimental Physics (NITEP), \\
        Osaka Metropolitan University, Osaka 558-8585, Japan} \\
        ${}^{c}$ \textit{KEK Theory Center, High Energy Accelerator Research Organization (KEK), \\
        Oho 1-1, Tsukuba, Ibaraki 305-0801, Japan} \\
        ${}^{d}$ \textit{The Graduate University for Advance Studies (SOKENDAI), \\
        Oho 1-1, Tsukuba, Ibaraki 305-0801, Japan} \\
    \end{center}
    \date{}
    \vspace*{5mm}
    \begin{abstract}
Grand gauge-Higgs unification of five dimensional SU(6) gauge theory on an orbifold $S^1/Z_2$ 
 with localized gauge kinetic terms is discussed. 
The Standard model (SM) fermions on the boundaries and 
 some massive bulk fermions coupling to the SM fermions on the boundary are introduced. 
Taking the power-law running contributions from the bulk fields into account, 
 perturbative gauge coupling unification is shown to be realized at around $10^{14}$ GeV, 
 which is a few order smaller than the unification scale of four dimensional grand unified theories. 
    \end{abstract}
\end{titlepage}
\section{Introduction}
Gauge-Higgs unification (GHU) \cite{1} is one of the physics beyond the Standard Model (SM),
	which solves the hierarchy problem
	by identifying the SM Higgs field with one of the extra spatial component of the higher dimensional gauge field.
In this scenario,
	the physical observables in Higgs sector are calculable and predictable regardless of its non-renormalizability.
For instance,
	the quantum corrections to Higgs mass and Higgs potential are known to be finite at one-loop \cite{2} and two-loop \cite{3}
	thanks to the higher dimensional gauge symmetry.

The hierarchy problem originally exists in grand unified theory (GUT)
	whether the discrepancy between the GUT scale and the weak scale are kept and stable under quantum corrections.
Therefore, the extension of GHU to grand unification is a natural direction to explore.
One of the authors discussed a grand gauge-Higgs unification (GGHU) \cite{4},
	\footnote{For earlier attempts and related recent works, see \cite{5}}
	where the five dimensional $SU(6)$ GGHU was considered
	and the SM fermions were embedded into zero modes of $SU(6)$ multiplets in the bulk.
This setup was very attractive because of the minimal matter content without massless exotic fermions absent in SM,
	namely an anomaly-free matter content.
However, 
 the down-type Yukawa couplings and the charged lepton Yukawa couplings in GHU 
 originated from the gauge interaction cannot be allowed 
 since the left-handed $SU(2)_L$ doublets and the right-handed $SU(2)_L$ singlets are embedded into different $SU(6)$ multiplets.
This fact seems to be generic in any GHU models as long as the SM fermions are embedded into the bulk fermions.
Fortunately, alternative approach to generate Yukawa coupling in a context of GHU has been known \cite{6,7},
	in which the SM fermions are introduced on the boundaries (i.e. fixed point in an orbifold compactification).
We also introduce massive bulk fermions, which couple to the SM fermions through the mass terms on the boundary.
Integrating out these massive bulk fermions leads to non-local SM fermion masses,
	which are proportional to the bulk to boundary couplings and exponentially sensitive to their bulk masses.
Then, the SM fermion mass hierarchy can be obtained by very mild tuning of bulk masses.

Along this line, we have improved an $SU(6)$ grand GHU model of \cite{4} in \cite{8},
	where the SM fermion mass hierarchy except for top quark mass was obtained 
	by introducing them on the boundary as $SU(5)$ multiplets,
	the four types of massive bulk fermions in $SU(6)$ multiplets coupling to the SM fermions.
Furthermore, we have shown that the electroweak symmetry breaking and an observed Higgs mass can be realized
	by introducing additional bulk fermions with large dimensional representation.
In GHU, generation of top quark mass is nontrivial 
	since Yukawa coupling is originally gauge coupling and fermion mass is at most an order of W boson mass as it stands.
As another known approach \cite{6},
	introducing the localized gauge kinetic terms on the boundary is known to have enhancement effects on fermion masses.
In a paper \cite{10}, we followed this approach in order to realize the SM fermion mass hierarchy including top quark.
Then, we have shown that the fermion mass hierarchy including top quark mass was indeed realized
	by appropriately choosing the bulk mass parameters and the size of the localized gauge kinetic terms.
The correct pattern of electroweak symmetry breaking was obtained by introducing extra bulk fermions as in our paper \cite{10},
	but their representations have become greatly simplified.

Next central issue is the gauge coupling unification, which should be explored in a context of GUT scenario.
It is well known that the gauge coupling running in (flat) large extra dimensions follows the power dependence
	on energy scale not logarithmic one \cite{11}.
Therefore, the GUT scale is likely to be very small comparing to the conventional 4D GUT.
It is therefore very nontrivial whether the unified $SU(6)$ gauge coupling at the GUT scale is perturbative
	since many bulk fields were introduced in our models \cite{8,10},
	which might lead to Landau pole below the GUT scale.
In fact, we saw that the perturbative gauge coupling unification cannot be realized
	because the number of the bulk fermions is too much in the previous setup \cite{10}.
Therefore, we had to reduce the number of the bulk fermions to avoid such a problem in our model discussed in our paper \cite{15}.
It was shown that this reduction will lead to additional generation mixings in the bulk.
Moreover, since we have changed now the bulk fermions couple to the SM fermions on the boundaries in our paper \cite{15},
	reproducing the SM fermion masses and generation mixings is nontrivial and their study should be reanalyzed.
We have shown that the SM fermion masses and mixing can be almost reproduced
	by mild tuning of bulk masses and the parameters of the localized gauge kinetic terms.
Our model \cite{15} was expected to overcome this issue.
In fact, we will see in this paper that the perturbative gauge coupling unification can be realized in our setup \cite{15}. 
The unification scale will be found to be around $10^{14}$ GeV, 
 which is a few order smaller than that of four dimensional grand unified theories.
Therefore, our model will be a good starting point for constructing a realistic model of GGHU.

This paper is organized as follows.
In the next section, we briefly review the gauge, Higgs and fermion sectors of our model.
In section 3, it is shown that the perturbative gauge coupling unification can be realized in our model.
Final section is devoted to our conclusions.
\section{Review of our model}
\subsection{Gauge and Higgs sector}
In this subsection, we briefly explain gauge and Higgs sectors of $SU(6)$ GHU model \cite{15}.
We consider a five dimensional (5D) $SU(6)$ gauge theory
	with an extra space compactified on an orbifold $S^1/Z_2$ with the radius $R$.
The orbifold has two fixed points at $y=0,\pi R$ where $y$ denotes the fifth coordinate
	and their $Z_2$ parities are given as follows.
  \begin{eqnarray}
		P  &=& \mbox{diag}(+,+,+,+,+,-) \, \, \mbox{at}~y=0, \nonumber \\
  	P' &=& \mbox{diag}(+,+,-,-,-,-) \, \, \mbox{at}~y=\pi R.
  \end{eqnarray}
The $Z_2$ parity for the gauge field and the scalar field originated from 
 an extra component of five dimensional gauge field are assigned
	as $A_{\mu}(-y) = PA_{\mu}(y)P^{\dag}$, $A_y(-y) = -PA_y(y)P^{\dag}$,
	which implies that $SU(6)$ gauge symmetry is broken to $SU(3)_C \times SU(2)_L \times U(1)_Y \times U(1)_X$
	by the combination of the symmetry breaking pattern at each boundary,
  \begin{eqnarray}
  	&& SU(6) \rightarrow SU(5)\times U(1)_X \hspace{1.6cm} \mbox{at}~y=0, \\
  	&& SU(6) \rightarrow SU(2)\times SU(4) \times U(1)' ~\mbox{at}~y=\pi R.
  \end{eqnarray}
The decomposition of the gauge field into the SM gauge group and their $\beta$ function are shown in Table \ref{Table1}
	which we will use for an analysis of the gauge coupling running in section 2.
The hypercharge $U(1)_Y$ is embedded in Georgi-Glashow $SU(5)$ GUT,
	where the weak mixing angle is $\sin^2{\theta_W}=3/8$ ($\theta_W$ : weak mixing angle) at the unification scale.

\begin{table}[b]
	\centering
  \begin{tabular}{|c|} \hline
  	gauge field $SU(6)\rightarrow SU(3)_C\times SU(2)_L\times U(1)_Y$ \\ \hline
    $35^{(+,+)} = (8,1)_{0}^{(+,+)}\oplus (1,3)_{0}^{(+,+)}\oplus (1,1)_{0}^{(+,+)}
									\oplus (1,1)_{0}^{(+,+)}$ \\
    							$\oplus (3,2)_{5/6}^{(+,-)} \oplus (3^*,2)_{-5/6}^{(+,-)} \oplus (3,1)_{-1/3}^{(-,+)}
									\oplus (3^*,1)_{1/3}^{(-,+)} \oplus (1,2)_{-1/2}^{(-,-)}\oplus (1,2)_{-1/2}^{(-,-)}$ \\ \hline \hline
    $\beta$ function ($b_3$, $b_2$, $b_1$) \\ \hline
    						$(3, 0, 0)+(0, 2, 0)+(0, 0, 0)+(0, 0, 0)$\\
    						$+(1, \frac{3}{2}, \frac{5}{2})+(1, \frac{3}{2}, \frac{5}{2})
								+(\frac{1}{2}, \frac{3}{2},\frac{1}{5})+(\frac{1}{2}, \frac{3}{2}, \frac{1}{5})
								+(0, \frac{1}{2}, \frac{3}{10})+(0, \frac{1}{2}, \frac{3}{10})$\\ \hline
	\end{tabular}
  \caption{Gauge field and its $\beta$ function.
						$r_{1,2}$ in $(r_1,~r_2)_a$ are $SU(3),~SU(2)$ representations in the SM, respectively.
						$a$ is $U(1)_Y$ charges.}
  \label{Table1}
\end{table}

The SM $SU(2)_L$ Higgs doublet field is identified with a part of an extra component of gauge field $A_y$ in a following,
	\begin{equation}
  	A_y = \frac{1}{\sqrt{2}}
    \left(
        	\begin{array}{c|c|c}
						\hspace{30pt} & \hspace{50pt} & H \\ \hline
            \hspace{30pt} & \hspace{50pt} & \hspace{30pt} \\
             && \\ \hline
             H^{\dag} & & \\
        	\end{array}
    \right).
	\end{equation}
We suppose that a vacuum expectation value (VEV) of the Higgs field is taken to be in the 28-th generator of $SU(6)$,
	$\langle A_y^a \rangle = \frac{2\alpha}{Rg}\delta^{a\,28}$,
	where $g$ is a 5D $SU(6)$ gauge coupling constant and $\alpha$ is a dimensionless constant.
The VEV of the Higgs field is given by $\langle H \rangle = \frac{\sqrt{2}\alpha}{Rg}$.
In this setup, the doublet-triplet splitting problem is solved by the orbifolding
	since the $Z_2$ parity of the colored Higgs is $(+,-)$ and it becomes massive \cite{12}.

After the Higgs field has the VEV,
	the mass eigenvalues of the 5D bulk fields are given by $m_n(q\alpha) = \frac{n+\nu+q\alpha}{R}$,
	where $n$ is KK mode, $\nu=0\,(1/2)$ is for a periodic (anti-periodic) boundary condition.
$q$ is an integer charge determined by the $SU(2)$ representation to which the field coupling to Higgs field belongs.
If the field with coupling to Higgs field belongs to $\textbf{N+1}$ representation of $SU(2)_L$,
	the integer charge $q$ is equal to $N$.

\subsection{Localized gauge kinetic term}
As mentioned in the introduction,
	we introduce additional localized gauge kinetic terms at $y=0$ and $y=\pi R$ to reproduce a top quark mass.
Lagrangian for $SU(6)$ gauge field is
	\begin{equation}
  	\mathcal{L}_g
			= -\frac{1}{4} \mathcal{F}^{a\,MN}\mathcal{F}^a_{MN}
				-2\pi R c_1 \delta(y)\frac{1}{4}\mathcal{F}^{b\,\mu\nu}\mathcal{F}^b_{\mu\nu}
				-2\pi R c_2 \delta (y-\pi R)\frac{1}{4}\mathcal{F}^{c\,\mu\nu}\mathcal{F}^c_{\mu\nu},
    \label{Equation10}
	\end{equation}
	where the first term is the bulk gauge kinetic term with 5D space-time indices $M,N=0,1,2,3,5$.
The second and the third terms are gauge kinetic terms with 4D space-time indices $\mu,\nu=0,1,2,3$ localized at fixed point.
$c_{1,2}$ are dimensionless free parameters.
The superscript $a,b,c$ denote the gauge indices for $SU(6)$, $SU(5)\times U(1)$, $SU(2)\times SU(4)\times U(1)'$.
Note that the localized gauge kinetic terms have only to be invariant under unbroken symmetries on each fixed point.

Because of the presence of localized gauge kinetic terms, the mass spectrum of the SM gauge field becomes very complicated.
In particular, their effects for a periodic sector and an anti-periodic sector are different,
	where the (anti-)periodic sector means the fields satisfying a condition $A(y+\pi R)=(-)A(y)$
	or the fields with parity $(P,P^{'})=(+,+),(-,-)((+,-),(-,+))$.
This difference originates from the boundary conditions for wave functions with a definite charge $q$, $f_n (y;q\alpha)$.
In a basis where 4D gauge kinetic terms are diagonal,
	we found them to be $f_n(y+\pi R;q\alpha)=e^{2i\pi q\alpha}f_n(y;q\alpha)$ in periodic sector
	and $f_n(y+\pi R;q\alpha)=e^{2i\pi(q\alpha +1/2)}f_n(y;q\alpha)$ in anti-periodic sector.
The KK mass spectrum of the SM gauge fields are obtained from the equation
	\begin{equation}
  	2(1-c_1 c_2 (\pi R m_n)^2)\sin^2(\pi R m_n)+(c_1+c_2)\pi R m_n\sin (2\pi R m_n) -2\sin^2(\pi(q\alpha+\nu))=0
  \end{equation}
	where $m_n$ is the KK mass.
The mass spectrum of the gauge field are deformed when $c = c_1 +c_2 \gg 1$.
In the case of introducing only one localized term $r \equiv c_1/(c_1+c_2) = 0$ or $1$,
	the mass spectrum tends to be shifted as follows. 
	\begin{equation}
		\begin{aligned}
			\frac{n+\nu+\alpha}{R} &\rightarrow \frac{n+\nu}{R}, \\
			\frac{n+\nu}{R}, \frac{n+\nu-\alpha}{R} &\rightarrow \frac{n+\nu-1/2}{R},
		\end{aligned}
		\label{massEq}
	\end{equation}
	where $\nu = 0$ or $1/2$.
Similarly, in the case of $r = c_1/(c_1+c_2) = 1/2$,
	the mass spectrum tends to be shifted, 
	\begin{equation}
		\begin{aligned}
			\frac{n+\nu+\alpha}{R} &\rightarrow \frac{n}{R}, \\
			\frac{n+\nu}{R}, \frac{n+\nu-\alpha}{R} &\rightarrow \frac{n-1}{R}.
		\end{aligned}
	\end{equation}

\subsection{Fermion sector}
In the paper \cite{10}, the SM fermions were embedded into $SU(5)$ multiplets localized at $y=0$ boundary,
	where three sets of decouplet, anti-quintet and singlet $\chi_{10}, \chi_{5^*}, \chi_1$ were introduced.
We also introduced three types of bulk fermions $\Psi$ and $\Tilde{\Psi}$ (referred as "mirror fermions")
	with opposite $Z_2$ parities each other per a generation and constant mass term
	such as $M\Bar{\Psi}\Tilde{\Psi}$ in the bulk to avoid exotic 4D massless fermions.
Without these mirror fermions and mass terms,
	we necessarily have extra exotic 4D massless fermions with the SM charges after an orbifold compactification.
In this setup, we have no massless chiral fermions the bulk and its mirror fermions.
The massless fermions are only the SM fermions and the gauge anomalies for the SM gauge groups are trivially canceled.

\begin{figure}[t]
	\centering
  \includegraphics[keepaspectratio, scale=0.7]{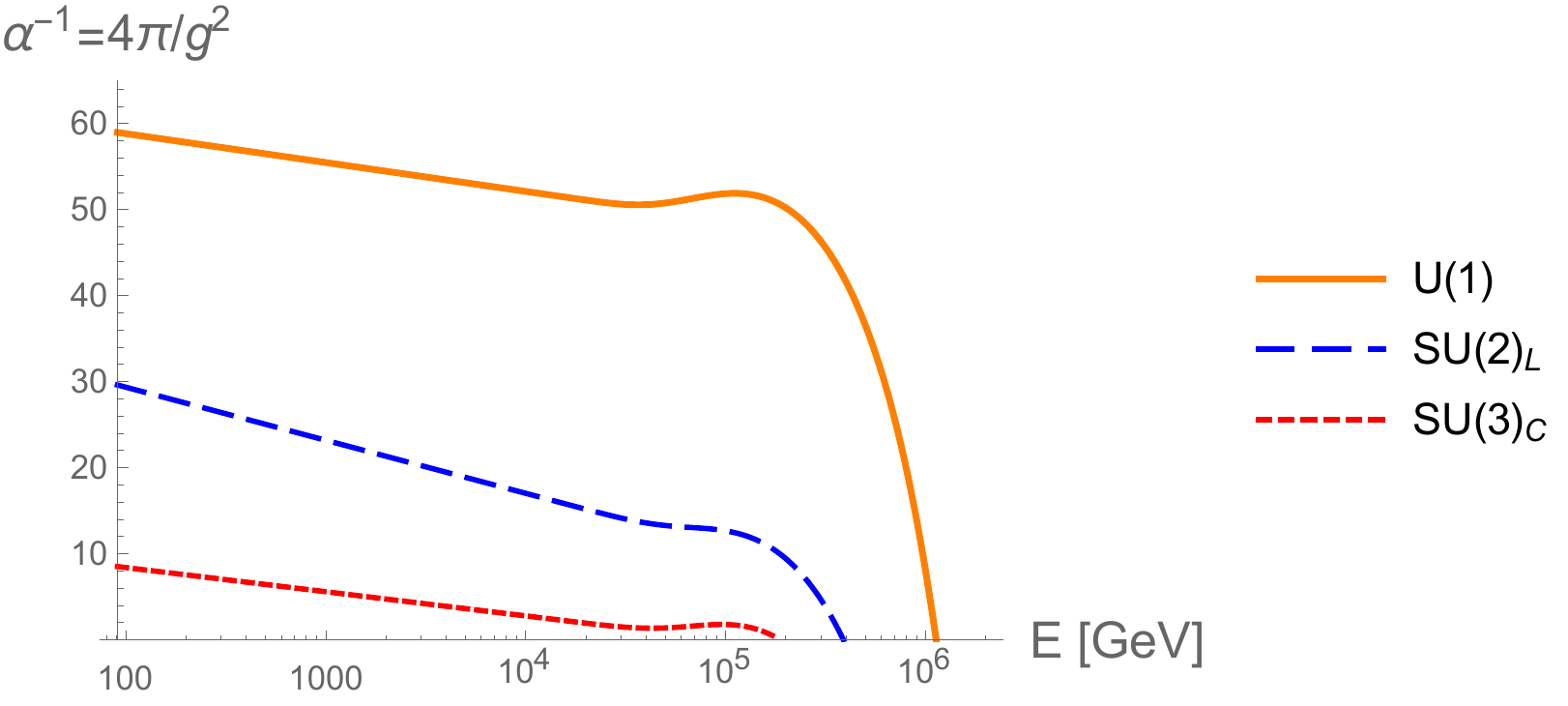}
  \caption{The SM gauge coupling running in our previous model \cite{10}.
					 The horizontal axis is an energy scale in GeV,
					 	and the verical axis is an inverse of the fine structure constant of the gauge coupling.
           The red dashed, blue dashed, and orange lines denote the running of $SU(3)_C$, $SU(2)_L$, and $U(1)_Y$ gauge coupling constants,
					 	respectively.
           The compactification scale is taken to be 10 TeV.}
  \label{Figure1}
\end{figure}

However, we could not obtain perturbative gauge coupling unification as shown in Fig.\,\ref{Figure1}
	because the number of the bulk fermions was too much in the previous setup as mentioned in the introduction.
As can be seen from Fig.\,\ref{Figure1}, all of the gauge couplings were not only unified,
	but also blown up around $10^{5\sim6}$ GeV,
	where the compactification scale is taken to be 10 TeV being a typical scale to realize 125 GeV Higgs mass in GHU scenario.
In order to avoid such a problem, we have succeeded in reducing the number of the bulk fermions in this model \cite{15},
	which also reproduces the fermion masses, mixing angles and a CP phase. From now on,
	fermion sector of our model will be shown briefly.
\footnote{In \cite{15}, we have shown the mechanism generating the SM masses
	and the generation mixings 
	based on this setup.}

In the setup of our model, we introduced five of bulk fermions $\Psi_{20},\Psi_{15},\Psi_{15'},\Psi_{6},\Psi_{6'}$
	and the corresponding mirror fermions shown in Table~\ref{Table2}.
\begin{table}[t]
	\centering
  \begin{tabular}{|c|c|} \hline
  	bulk fermion $SU(6) \rightarrow SU(5)$ & mirror fermion \\ \hline
    $20^{(+,+)}=10\oplus10^*$ & $20^{(-,-)}$ \\ \hline
    $15^{(+,+)}=10\oplus5$ & $15^{(-,-)}$ \\ \hline
    $15'^{(+,-)}=10'\oplus5'$ & $15'^{(-,+)}$ \\ \hline
    $6^{(-,-)}=5\oplus1$ & $6^{(+,+)}$ \\ \hline
    $6'^{(+,+)}=5'\oplus1'$ & $6'^{(-,-)}$ \\ \hline
  \end{tabular}
	\vspace{-5pt}
  \caption{Representation of bulk fermions and the corresponding mirror fermions.
					 $R$ in $R^{(+,+)}$ means an $SU(6)$ representation of the bulk fermion.
					 $r_i$ in $r_1\oplus r_2$ are $SU(5)$ representations.}
  \label{Table2}
\end{table}
The SM quarks and leptons for the first and the second generation were embedded into $SU(5)$ multiplets
	localized at $y=0$ boundary, which were two sets of decouplet, anti-quintet and singlet $\chi_{10},\, \chi_{5^*},\, \chi_1$.
On the other hand,
	those for the third generation were embedded into $SU(3)_C\times SU(2)_L\times U(1)_Y$ multiplets
	localized at $y=\pi R$ boundary.
\begin{table}[t]
	\centering
  \begin{tabular}{|c|c|} \hline
  	bulk fermion $SU(5)\rightarrow SU(3)_C\times SU(2)_L\times U(1)_Y$
				& $\beta$ function ($\Tilde{b}_3$, $\Tilde{b}_2$, $\Tilde{b}_1$) \\ \hline
    $10 = Q_{20}(3,2)_{1/6}^{(+,+)}\oplus U^*_{20}(3^*,1)_{-2/3}^{(+,-)}\oplus E^*_{20}(1,1)_{1}^{(+,-)}$
				& (1, $\frac{3}{2}$, $\frac{1}{10}$), ($\frac{1}{2}$, 0, $\frac{4}{5}$), (0, 0, $\frac{3}{5}$)\\ \hline
    $10^* = Q^*_{20}(3^*,2)_{-1/6}^{(-,-)}\oplus U_{20}(3,1)_{2/3}^{(-,+)}\oplus E_{20}(1,1)_{-1}^{(-,+)}$
				& (1, $\frac{3}{2}$, $\frac{1}{10}$), ($\frac{1}{2}$, 0, $\frac{4}{5}$), (0, 0, $\frac{3}{5}$)\\ \hline
  \end{tabular}
  \vspace{-5pt}
  \caption{$\textbf{20}$ bulk fermion and their $\beta$ function.
					 $r_{1,2}$ in $(r_1,~r_2)_a$ are $SU(3),~SU(2)$ representations in the SM, respectively.
					 $a$ is $U(1)_Y$ charges.}
  \label{Table3}
	\vspace{15pt}
  \centering
  \begin{tabular}{|c|c|} \hline
  	bulk fermion $SU(5)\rightarrow SU(3)_C\times SU(2)_L\times U(1)_Y$
				& $\beta$ function ($\Tilde{b}_3$, $\Tilde{b}_2$, $\Tilde{b}_1$) \\ \hline
    $10 = Q_{15}(3,2)_{1/6}^{(+,-)}\oplus U^*_{15}(3^*,1)_{-2/3}^{(+,+)}\oplus E^*_{15}(1,1)_{1}^{(+,+)}$
				& (1, $\frac{3}{2}$, $\frac{1}{10}$), ($\frac{1}{2}$, 0, $\frac{4}{5}$), (0, 0, $\frac{3}{5}$)\\ \hline
    $5 = D_{15}(3,1)_{-1/3}^{(-,+)}\oplus L^*_{15}(1,2)_{1/2}^{(-,-)}$
				& ($\frac{1}{2}$, 0, $\frac{1}{5}$), (0, $\frac{1}{2}$, $\frac{3}{10}$)\\ \hline \hline
    bulk fermion $SU(5)\rightarrow SU(3)_C\times SU(2)_L\times U(1)_Y$
				& $\beta$ function ($\Tilde{b}_3$, $\Tilde{b}_2$, $\Tilde{b}_1$) \\ \hline
    $10' = Q_{15'}(3,2)_{1/6}^{(+,+)}\oplus U^*_{15'}(3^*,1)_{-2/3}^{(+,-)}\oplus E^*_{15'}(1,1)_{1}^{(+,-)}$
				& (1, $\frac{3}{2}$, $\frac{1}{10}$), ($\frac{1}{2}$, 0, $\frac{4}{5}$), (0, 0, $\frac{3}{5}$)\\ \hline
    $5' = D_{15'}(3,1)_{-1/3}^{(-,-)}\oplus L^*_{15'}(1,2)_{1/2}^{(-,+)}$
				& ($\frac{1}{2}$, 0, $\frac{1}{5}$), (0, $\frac{1}{2}$, $\frac{3}{10}$)\\ \hline
  \end{tabular}
  \vspace{-5pt}
  \caption{Upper (Lower) table shows $\textbf{15}$ ($\textbf{15}'$) bulk fermion and their $\beta$ function.
					 $r_{1,2}$ in $(r_1,~r_2)_a$ are $SU(3),~SU(2)$ representations in the SM, respectively.
					 $a$ is $U(1)_Y$ charges.}
  \label{Table4}
	\vspace{15pt}
  \centering
  \begin{tabular}{|c|c|} \hline
  	bulk fermion $SU(5)\rightarrow SU(3)_C\times SU(2)_L\times U(1)_Y$
				& $\beta$ function ($\Tilde{b}_3$, $\Tilde{b}_2$, $\Tilde{b}_1$) \\ \hline
    $5 = D_{6}(3,1)_{-1/3}^{(-,+)}\oplus L^*_{6}(1,2)_{1/2}^{(-,-)}$
				& ($\frac{1}{2}$, 0, $\frac{1}{5}$), (0, $\frac{1}{2}$, $\frac{3}{10}$) \\ \hline
    $1 = N^*_{6}(1,1)_{0}^{(+,+)}$ & (0, 0, 0) \\ \hline \hline
    bulk fermion $SU(5)\rightarrow SU(3)_C\times SU(2)_L\times U(1)_Y$
				& $\beta$ function ($\Tilde{b}_3$, $\Tilde{b}_2$, $\Tilde{b}_1$) \\ \hline
    $5' = D_{6'}(3,1)_{-1/3}^{(-,-)}\oplus L^*_{6'}(1,2)_{1/2}^{(-,+)}$
				& ($\frac{1}{2}$, 0, $\frac{1}{5}$), (0, $\frac{1}{2}$, $\frac{3}{10}$) \\ \hline
    $1' = N^*_{6'}(1,1)_{0}^{(+,-)}$
				& (0, 0, 0)\\ \hline
  \end{tabular}
  \vspace{-5pt}
  \caption{Upper (Lower) table shows $\textbf{6}$ ($\textbf{6}'$) bulk fermion and their $\beta$ function.
					 $r_{1,2}$ in $(r_1,~r_2)_a$ are $SU(3),~SU(2)$ representations in the SM, respectively.
					 $a$ is $U(1)_Y$ charges.}
  \label{Table5}
\end{table}
The decomposition of the introduced bulk fermions
	in the $\textbf{20}$, $\textbf{15}~(\textbf{15}')$, $\textbf{6}~(\textbf{6}')$ representations
	into the SM gauge group and their $\beta$ function are summarized
	in Tables \ref{Table3}, \ref{Table4}, \ref{Table5}, respectively.

Solving the exact KK spectrum of the bulk fermions from this Lagrangian is a very hard task
	because of the complicated bulk and boundary system.
We assume in this paper that the physical mass induced for the boundary fields is much smaller than the masses of the bulk fields \cite{6}.
This is reasonable since the compactfication scale and
	the bulk mass mainly determining the KK mass spectrum of the bulk fields is larger than the mass for the boundary fields whose typical scale is given by the Higgs VEV.
In this case, the effects of the mixing on the spectrum for the bulk fields can be negligible and
	the spectrum $m_n^2 = (\frac{\lambda}{\pi R})^2+m_n(q\alpha)^2$ is a good approximation \cite{6}.

\section{Gauge coupling unification}
In ordinary 4D field theories, the gauge couplings $g_i~(i=1,2,3)$ of the Standard Model are dimensionless.
They evolve as the following one-loop renormalization group equation (RGE)
  \begin{equation}
    \frac{d}{d\ln{\mu}} \alpha_i^{-1}(\mu) = - \frac{b_i}{2\pi},
  \end{equation}
  whose solution is given by
  \begin{equation}
    \alpha_i^{-1}(\mu) = \alpha_i^{-1}(M_Z) - \frac{b_i}{2\pi}\ln{\frac{\mu}{M_Z}}.
  \end{equation}
This is the usual logarithmic running of the gauge couplings.
Here $\alpha_i \equiv g_i^2/4\pi$, the $b_i$ are the one-loop beta-function coefficients for the Standard Model gauge group
  \begin{equation}
    (b_1, b_2, b_3) = (41/10, -19/6, -7)
  \end{equation}
  and we have taken the Z-mass $M_Z\equiv$ 91.17 GeV as an arbitrary low-energy reference scale.

These gauge couplings also receive corrections in extra dimensions,
 and we can calculate such corrections in the usual way by evaluating the same one-loop diagrams
 (particularly the vacuum polarization diagram) as shown in \cite{11}.
The full one-loop corrected gauge coupling is given as
 \begin{equation}
     \alpha_i^{-1}(\Lambda)
      = \alpha_i^{-1}(\mu)
        - \frac{\Tilde{b}_i}{4\pi} \int^{\frac{\pi}{4}\mu^{-2}}_{\frac{\pi}{4}\Lambda^{-2}}\frac{dt}{t}\mathcal{P}(t)
     \label{Equation32}
 \end{equation}
 where $\Tilde{b}_i$ are new beta-function coefficients by bulk fermion contributions. 
 $\mu$ is a renormalization scale and $\Lambda$ is the cutoff scale of 5D theory.  
 $\mathcal{P}(t)$ denotes the contribution from the bulk fields with KK mass spectrum $m_n$,
 \begin{equation}
    \mathcal{P}(t) \equiv \sum^{\infty}_{n=-\infty} \exp{\left\{-tm_n^2\right\}}.
 \end{equation}

\begin{table}[t]
  \centering
  \begin{tabular}{|c|c|} \hline
    KK mass $m_n^2$ & $\mathcal{P}(t)$\\ \hline
    $r_0^2\delta_{n,0}/R^2$ & $\exp[-r_0^2t/R^2]$ \\ \hline
    $\{(n+\alpha)^2+\lambda^2\}/R^2$ & $\theta_3(i\alpha t/R^2,\exp[-t/R^2])\exp[-t(\alpha^2+\lambda^2)/R^2]$\\ \hline
    $\{(n+1/2+\alpha)^2+\lambda^2\}/R^2$ & $\theta_2(i\alpha t/R^2,\exp[-t/R^2])\exp[-t(\alpha^2+\lambda^2)/R^2]$ \\ \hline
  \end{tabular}
  \caption{The results of evaluation of $\mathcal{P}(t)$ depending on the KK mass.}
  \label{Table6}
\end{table}

\begin{figure}[t]
  \begin{tabular}{cc}
  \begin{minipage}[c]{0.47\hsize}
  \centering
  \includegraphics[keepaspectratio, scale=0.8]{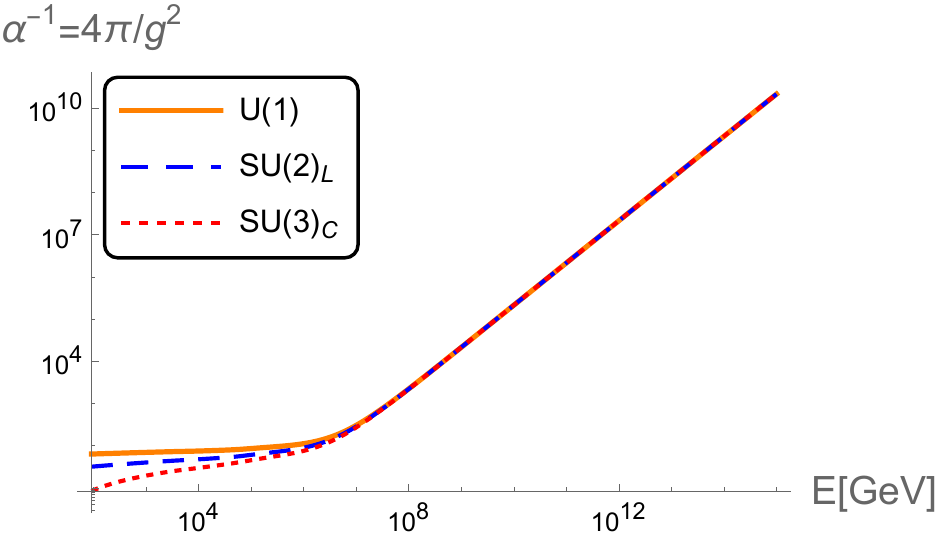}
  \end{minipage}&
  \begin{minipage}[c]{0.47\hsize}
  \centering
  \includegraphics[keepaspectratio, scale=0.8]{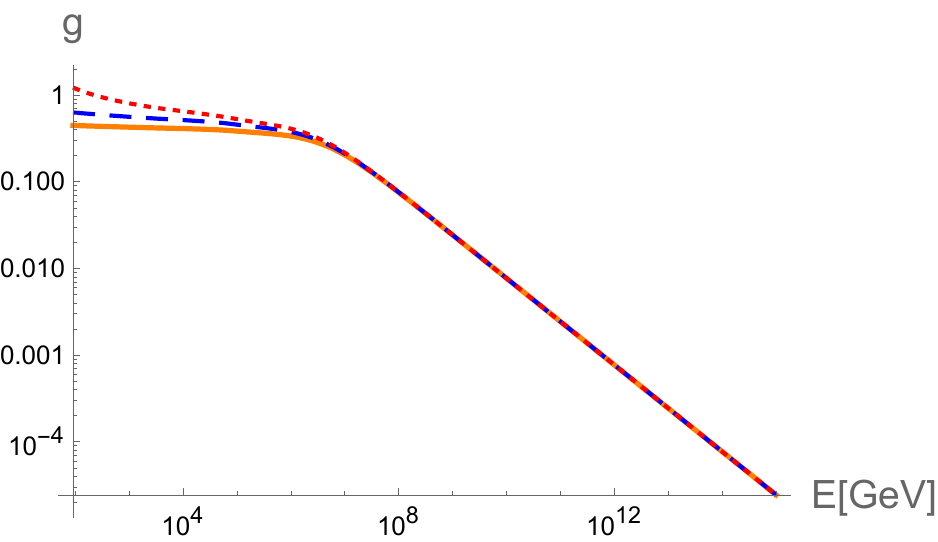}
  \end{minipage}\\
  &\\
  &\\
  \begin{minipage}[c]{0.47\hsize}
  \centering
  \includegraphics[keepaspectratio, scale=0.8]{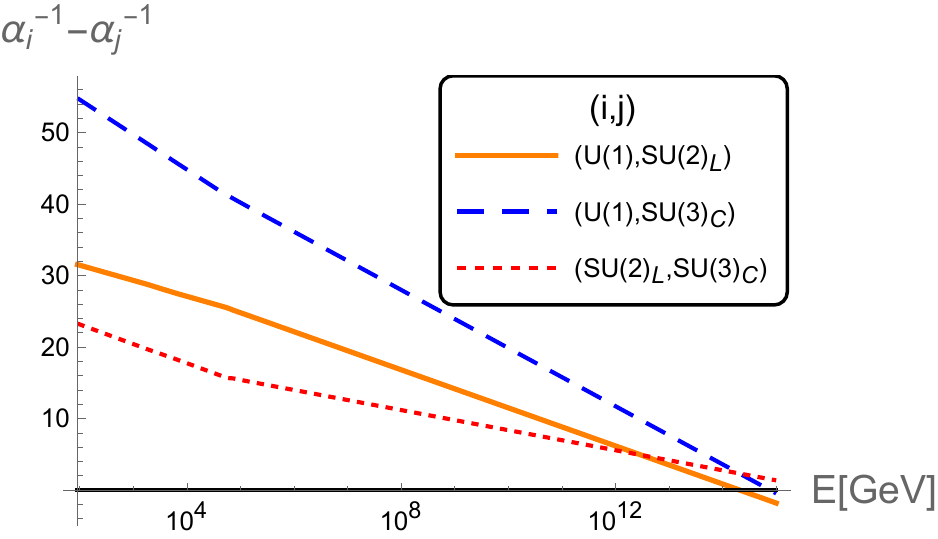}
  \end{minipage}&
  \begin{minipage}[c]{0.47\hsize}
  \centering
  \includegraphics[keepaspectratio, scale=0.8]{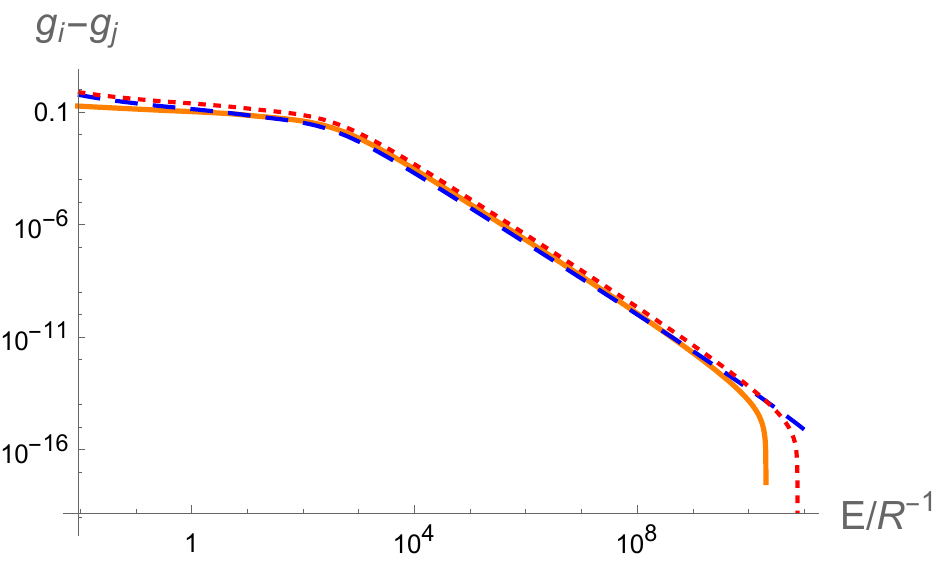}
  \end{minipage}
  \end{tabular}
  \caption{The perturbative gauge coupling unification
            in the case of $c=80$, $r=0$ and $R^{-1}=10$ TeV.
           The upper figures show the energy dependence of gauge coupling
            $\alpha^{-1}$ (left) and $g$ (right).
           The lower figures show the energy dependence of differences between each pair of the gauge couplings
           $\alpha_i^{-1}-\alpha_j^{-1}$ (left) and $g_i-g_j$ (right).}
  \label{Figure2}
 \end{figure}

The results are expressed by the elliptic theta functions $\theta_i~(i=2,3)$
  \begin{eqnarray}
      \theta_2(v,q) &=& 2q^{1/4}\sum^{\infty}_{n=0}q^{n(n+1)}\cos[(2n+1)v], \\
      \theta_3(v,q) &=& 1 + 2\sum^{\infty}_{n=0}q^{n^2}\cos[2nv].
  \end{eqnarray}
Note that the results are different depending on whether the fields are periodic or anti-periodic as can be seen from Table \ref{Table6}.
This means that we should be careful for the periodicity of the fields listed
 in Table \ref{Table1}, \ref{Table3}, \ref{Table4}, \ref{Table5}, \ref{Table6} in our analysis. 
%
The elliptic theta functions $\theta_i~(i=2,3)$ can be approximated to
\begin{equation}
  \theta_i(0,\exp[-t/R^2]) \sim R\sqrt{\frac{\pi}{t}}
  \label{Equation33}
\end{equation}
in the case of $t/R^2\ll 1$ which corresponds to the assumption that both $\mu$ and $\Lambda$ are much larger than $R^{-1}$.
After substituting this approximation (\ref{Equation33}) into (\ref{Equation32}) and evaluating the integral over $t$,
  RGE (\ref{Equation32}) becomes the following expression
\begin{equation}
  \alpha_i^{-1}(\Lambda)
    = \alpha_i^{-1}(\mu)
      - \frac{b_i-\Tilde{b}_i^{(+)}}{4\pi}\ln \frac{\Lambda}{\mu}
      - \frac{\Tilde{b}_i^{(+)}+\Tilde{b}_i^{(-)}}{\pi}R(\Lambda-\mu).
      \label{alpha}
\end{equation}
The third term on the right-hand side shows the power-law dependence of the gauge coupling on the energy scale.
Here $+\,(-)$ in  $\Tilde{b}_i^{(+(-))}$ shows that its contribution comes from the (anti-)periodic fields.
Note that the bulk fermion and the corresponding mirror fermion have the same $\beta$ function because they have the same periodicity.
Asymptotic freedom of gauge couplings can be confirmed
  by the fact that the beta function for KK mode $\Tilde{b}^{(+)}+\Tilde{b}^{(-)}$ is negative.
It can be calculated by using informations in Table \ref{Table1}, \ref{Table3}, \ref{Table4}, \ref{Table5}, 
  \begin{equation}
    \Tilde{b}_i^{(+)}+\Tilde{b}_i^{(-)} = -\frac{2}{3} < 0.
    \label{betas}
  \end{equation}
Therefore, the perturbative gauge coupling unification is expected 
 in this model, which cannot be realized in the previous model \cite{10} 
 due to the large number of the introduced bulk fermions.
Fig.\,\ref{Figure2} shows
  energy dependences of the gauge couplings and differences between each pair of gauge couplings
  at $c=80$, $r=0$ and $R^{-1}=10$ TeV.
In this case,
  the unification scale $M_{G}$ and unification coupling $\alpha_{G}^{-1}(g_{G})$ are identified with 
  the scale where $U(1)$ and $SU(2)_{L}$ couplings are unified and we obtain 
  $M_{G} \sim 2.1 \times 10^{14}$ GeV and
  $\alpha_{G}^{-1} \sim 4.4 \times 10^{9} $,
  ($g_{G} \sim 5.3 \times 10^{-5}$).
The difference between the unification coupling and $SU(3)_{C}$ coupling at $M_{G}$ is
  $\left|(\alpha_{G}^{-1} -\alpha_3^{-1})/\alpha_{G}^{-1}\right|  \sim 5 \times 10^{-10} $
  ($\left|(g_{G} -g_3)/g_{G}\right| \sim 2.6 \times 10^{-10} $),
  therefore three gauge couplings unify with an accuracy of $10^{-10}$.
Alternatively, assuming the unification of three couplings $\alpha_G$ at $M_{G}$,
 and evolving $SU(3)$ coupling down to the weak scale by RGE, 
  $\alpha_{3}^{-1}(M_z) \sim 10.7$ ($g_{3} \sim 1.08$) is found,
  which is larger (smaller) than the experimental value $\alpha_3 \sim 8.4$
  ($g_3 \sim 1.2$).
We also analyze $r=1/2$ and $r=1$ cases.
In the former case,
  almost the same result as $r=0$ case is obtained.
In the latter case,
  the differences are smaller ($\sim 10^{-11}$),
  unification scale is larger ($M_{G} \sim 4.1 \times 10^{15}$)
  and $SU(3)$ coupling at the weak scale is larger ($\alpha_3 \sim 13.2$).
We analyze the coupling unification in other parameter cases,
  $(c,r,R^{-1})$ = $(80,0,15\mbox{ TeV})$, $(90,0,10\mbox{ TeV})$, $(90,0,15\mbox{ TeV})$ shown in previous paper \cite{15}
  and the results are shown in Table~\ref{Table7}.
The unification scale in our model is comparable to that of four-dimensional GUT,
  since the running of coupling constant in $t/R^{2} \gg 1$ region is dominated
  by the contributions linearly dependent on the energy scale as in Eq.~(\ref{alpha})
  and their beta functions (Eq.~\ref{betas}) are common,
  then the differences between each pair of the gauge couplings are dominated by the logarithmic terms.
In each case shown in Table~\ref{Table7},
   the theoretical value for $SU(3)$ coupling with one-loop corrections at the weak scale
   are slightly deviated from the experimental value.
However,
  it is possible that we could analyze two-loop corrections to obtain
  more accurate unification since the difference between the each pair of the couplings at $M_G$ is extremely small. 

  \begin{table}[t]
    \centering
    \begin{tabular}{|c|c|c||c|c|c|c|} \hline
      $c$ & $r$ & $R^{-1}$& $M_{G}$ & $\alpha_{G}^{-1}$
                      & $\left|(\alpha_{G}^{-1} -\alpha_3^{-1})/\alpha_{G}^{-1}\right|$
                      &  $\alpha^{-1}_{3}(M_{Z})$ \\ \hline
      $80$ & 0 &  $10$ TeV
                      & $2.1 \times 10^{14}$ GeV
                      & $4.4 \times10^{9}$
                      & $5.26 \times 10^{-10}$
                      & $10.7$ \\
      $80$ & 0 & $15$ TeV
                      & $2.2 \times 10^{14}$ GeV
                      & $3.2 \times 10^{10}$
                      & $6.12 \times 10^{-10}$
                      & $10.4$\\
      $90$ & 0 & $10$ TeV
                      & $2.1 \times 10^{14}$ GeV
                      & $4.3 \times 10^{9}$
                      & $5.25 \times 10^{-10}$
                      & $10.7$ \\
      $90$ & 0 & $15$ TeV
                      & $2.3 \times 10^{14}$ GeV
                      & $3.2 \times 10^{9}$
                      & $6.1 \times 10^{-10}$
                      & $10.4$  \\ \hline
    \end{tabular}
    \caption{The results of gauge coupling unification analysis at $r=0$.
             The unification scale $M_G$ and the unification coupling $\alpha_G^{-1}$ are identified with
             the scale where $U(1)$ and $SU(2)_L$ couplings are unified.
             $\left|(\alpha_{G}^{-1} -\alpha_3^{-1})/\alpha_{G}^{-1}\right|$ is
              the difference between $\alpha_{G}^{-1}$ and $SU(3)_C$ coupling at $M_G$.
             $\alpha_3^{-1}(M_Z)$ is the $SU(3)$ coupling at weak scale,
              assuming that three gauge couplings are unified to $\alpha_G^{-1}$ at $M_G$.
            }
    \label{Table7}
  \end{table}

We further analyze the coupling unification in the case of larger compactification scale, 
 which was not analyzed in the previous paper \cite{15}.
Although this case is not realistic since the Higgs mass is likely to be enhanced, 
 the SM fermion masses and mixings can be reproduced in larger compactification scale.
The results are shown in Table~\ref{Table8}. 
In the range of $R^{-1}= 200$ TeV $-$ $220$ TeV,
  $SU(3)$ coupling at the weak scale can be within the error range of
  experimental value.

\begin{table}[t]
  \centering
  \begin{tabular}{|c|c|c||c|c|c|} \hline
    $c$ & $r$ & $R^{-1}$  & $M_{G}$ &  $\alpha_{G}^{-1}$
                    & $\alpha_{3}^{-1}(M_{Z})$ \\ \hline
    $80$ & 0 & $200$ TeV
                     & $3.8 \times 10^{14}$ GeV
                     & $4.1 \times 10^{8}$
                     & $8.55$ \\
    $80$ & 0 & $220$ TeV
                     & $4.0 \times 10^{14}$ GeV
                     & $3.8 \times 10^{8}$
                     & $8.49$ \\ \hline
  \end{tabular}
  \caption{The results of gauge coupling unification analysis
            in the case of large compactified scale at $r=0$.
            The unification scale $M_G$ and the unification coupling $\alpha_G^{-1}$ are identified with
            the scale where $U(1)$ and $SU(2)_L$ couplings are unified.
            $\alpha_3^{-1}(M_Z)$ is the $SU(3)$ coupling at the weak scale,
             assuming that three gauge couplings are unified to $\alpha_G^{-1}$ at $M_G$. }
  \label{Table8}
\end{table}

\section{Conclusions}
In this paper, we have discussed $SU(6)$ GGHU with localized gauge kinetic terms.
The SM fermions are introduced on the boundaries.
We also introduced massive bulk fermions in three types of $SU(6)$ representations coupling to the SM fermions on the boundaries.
The number of them has been reduced in order to achieve perturbative gauge coupling unification which could not be realized in \cite{8,10}.
It was shown in this paper that the perturbative gauge coupling unification can be indeed realized in our model \cite{15}. 
Remarkably, the unification scale in our model was found to be $10^{14}$ GeV, 
 which is a few order smaller than the 4D GUT scale $10^{15-16}$ GeV. 
This is because the beta functions for introduced bulk fermions are common to each gauge coupling running 
 and the differences between each pair of gauge couplings are dominated by the logarithmic contributions in RGE. 
Our model turned out to be indeed a good starting point for constructing a realistic model of GGHU.

There is an issue to be explored in a context of GUT scenario, namely, a proton decay.
In large extra dimension models such as GHU discussed in this paper,
  $X, Y$ gauge boson masses are likely to be
  light comparing to the conventional GUT scale due to the power law running of the gauge coupling.
Therefore, proton decays very rapidly and our model is immediately excluded 
 by the experimental constraints from the Super Kamiokande data as it stands.
Possible dangerous baryon number violating operators must be forbidden,
  for instance, by some symmetry (see \cite{14} for UED case) for the proton stability.
If $U(1)_X$ in our model is broken to some discrete symmetry which plays its role, it would be very interesting.
It would be also interesting to investigate the main decay mode of the proton decay in our model 
and give predictions for Hyper Kamiokande experiments.
\subsection*{Acknowledgments}
This work was supported by JST SPRING, Grant Number JPMJSP2139 (YY).


\end{document}